\documentclass[epjCONF,columns]{svjour} 
\usepackage{graphics}
\usepackage[varg]{txfonts} 
\usepackage[latin1]{inputenc}
\session-title{NSRT12}
\begin{document}
\title{Effects of phonon-phonon coupling on properties of pygmy
resonance in $^{124-132}$Sn}
\author{N. N. Arsenyev\inst{1}\fnmsep\thanks{\email{arsenev@theor.jinr.ru}}
\and A. P. Severyukhin\inst{1}\and V. V. Voronov\inst{1} \and
Nguyen Van Giai\inst{2}}

\institute{Bogoliubov Laboratory of Theoretical Physics, Joint
Institute for Nuclear Research, 141980 Dubna, Moscow region,
Russia \and Institut de Physique Nucl\'eaire, CNRS-IN2P3,
Universit\'e Paris-Sud, F-91406 Orsay Cedex, France}

\abstract{ Starting from an effective Skyrme interaction we study
effects of phonon-phonon coupling on the low-energy electric
dipole response in $^{124-132}$Sn. The QRPA calculations are
performed within a finite rank separable approximation. The
inclusion of two-phonon configurations gives a considerable
contribution to low-lying strength. Comparison with available
experimental data shows a reasonable agreement for the low-energy
$E1$ strength distribution. }

\maketitle

\section{Introduction}
Exotic nuclear collective excitations, like the pygmy dipole
resonance, represent a subject of intense investigations du\-ring
last decades, see for example \cite{gbp98,akfb05,kpaf07,oe07}. The
structure and dynamics of low-energy dipole strength, also
referred to as pygmy dipole resonance (PDR), has extensively been
investigated using a variety of theoretical approaches and models
\cite{pvkc07}. Recent studies have made use of the Hartree-Fock
(HF) plus random phase approximation (RPA) \cite{as10}, the
Hartree-Fock-Bogoliubov (HFB) model plus quasiparticle RPA (QRPA)
\cite{tsnv07,agkk11}, QRPA plus phonon coupling \cite{sbc04}, the
quasiparticle phonon model (QPM) including complex configurations
\cite{gbp98,tl08}, the relativistic RPA \cite{pie06} and QRPA
\cite{lrt07}. Also the quasiparticle time blocking approximation
(QTBA) has been used either in a non-relativistic framework
\cite{agkk11} or with relativistic Lagrangians (RQTBA)
\cite{lrt08,lrtl09}.

One of the successful tools for descri\-bing the PDR is the QRPA
with the self-consistent mean-field derived from Skyrme effective
nucleon-nucleon interactions. Such an approach describes the
properties of the low-lying states less accurately than more
phenomenological ones, but the results are in a reasonable
agreement with experimental data (Ref. \cite{pvkc07} and
references therein). Due to the anharmonicity of vibrations there
is a coupling between one-phonon and more complex states
\cite{bm75,solo}. The main difficulty is that the complexity of
calculations beyond standard QRPA increases rapidly with the size
of the configuration space, so one has to work within limited
spaces. Using a finite rank separable approximation suggested in
\cite{gsv98,svg08,svg04} for the residual interaction resulting
from Skyrme forces one can overcome this difficulty. In this paper
we study the properties of the low-lying and high-lying electric
dipole strength in the neutron-rich Sn isotopes. The couplings
between one- and two-phonon components in the wave functions of
excited states are taken into account.

\section{The method}
The calculations are performed by using the SLy4 \cite{sly4}
interaction in the particle-hole (\rm{p-h}) channel and a
density-dependent zero-range interaction in the particle-particle
(\rm{p-p}) channel. Spherical symmetry is assumed for the
HF\linebreak ground states. The strength of the surface-peaked
zero-ran\-ge pairing force is taken equal to -940 MeV~fm$^3$ in
connection with the soft cutoff at 10 MeV above the Fermi energy
as introduced in Ref. \cite{svg08}. This value of the pairing
strength is fitted to reproduce the experimental pairing energies
for both protons and neutrons.

The residual interaction in the \rm{p-h} channel $V^{\rm ph}_{\rm
res}$ and in the \rm{p-p} channel $V^{\rm pp}_{\rm res}$ can be
obtained as the second derivative of the energy density functional
with respect to the particle density $\rho$ and the pair density
$\tilde{\rho}$, respectively. Following Ref. \cite{gsv98} we
simplify $V^{\rm ph}_{\rm res}$ by approximating it by its
Landau-Migdal form. Moreover we neglect the $l=1$ Landau
parameters (Landau parameters with $l> 1$ are equal to zero in the
case of Skyrme interactions). In this work we study only normal
parity states and one can neglect the spin-spin terms since they
play a minor role. The two-body Coulomb and spin-orbit residual
interactions are also dropped. The expressions for $F^{\rm ph}_0,
F^{\rm 'ph}_0$ and $F^{\rm pp}_0, F^{\rm 'pp}_0$ can be found in
Ref. \cite{sg81} and in Ref. \cite{svg08}, respectively. The
Landau parameters $F_0$, $G_0$, $F^{'}_0$, $G^{'}_0$ expressed in
terms of the Skyrme force parameters \cite{sg81} depend on $k_{\rm
F}$. As is pointed out in our previous works \cite{gsv98,svg08}
one needs to adopt some effective value for $k_{\rm F}$ to give an
accurate representation of the original \rm{p-h} Skyrme
interaction. For the present calculations we use the nuclear
matter value for $k_{\rm F}$.

In describing the giant dipole resonance (GDR) we sho\-uld exclude
the spurious state due to the center-of-mass motion of the
nucleus. The spurious state admixture can be present as a
component in each of the wave functions of the excited states. The
spurious $1^-$ state is excluded from the excitation spectra by
introducing the effective neutron $q_{\rm n}= -Z/A$ and proton
$q_{\rm p}= N/A$ charges \cite{bm75}. An alternative way of
eliminating the spurious state is to orthogona\-li\-ze it with
respect to all physical states \cite{c00}. At was shown
\cite{as10} that eliminating the spurious state by means of
effective charges or orthogona\-li\-zing it to all physical states
lead to very similar results.

We take into account the coupling between the one- and two-phonon
components in the wave functions of excited states. Thus, in the
simplest case one can write the wave functions of excited states
as \cite{solo}
\begin{eqnarray}
\Psi _\nu (JM) = \left(\sum_iR_i(J\nu)Q_{JMi}^{+}\right.
\nonumber\\
\left.+\sum_{\lambda _1i_1\lambda _2i_2}P_{\lambda _2i_2}^{\lambda
_1i_1}(J\nu)\left[ Q_{\lambda _1\mu _1i_1}^{+}Q_{\lambda _2\mu
_2i_2}^{+}\right] _{JM}\right)|0\rangle. \label{wf2ph}
\end{eqnarray}
where $\mid 0\rangle$ is the phonon vacuum, $Q_{\lambda \mu
i}^{+}\mid0\rangle$ is the phonon creation operator and $\nu$
labels the excited states. The coefficients $R_i(J\nu)$,
$P_{\lambda_2i_2}^{\lambda_1i_1}(J\nu)$ and energies of the
excited states $E_{\nu}$ are determined by solving the
corresponding se\-cu\-lar equation of Ref. \cite{svg04}.

The two-phonon configurations of the wave function (\ref{wf2ph})
are constructed from natural-parity phonons with multipolarities
$\lambda=1,2,3,4,5$. All dipole excitations with ener\-gies below
35 MeV and 15 most collective phonons of $\lambda=2,3,4,5$
multipolarity are included in the wave function (\ref{wf2ph}). It
is found that extending the model space for one-phonon
configurations does not change much the calculated energies and
transition probabilities.

\section{Results and discussion}
\begin{figure}
\resizebox{1.0\columnwidth}{!}{\includegraphics{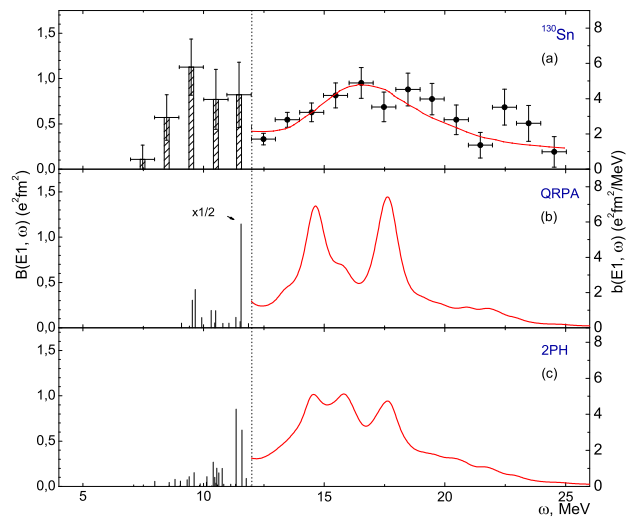}}
\caption{$B(E1)$ strength distributions in $^{130}$Sn. (a)
Experimental data \cite{akfb05,kpaf07}; (b) QRPA results; (c) QRPA
plus phonon-phonon coupling results.}
\label{Sn130}       
\end{figure}

In Figure \ref{Sn130} the calculated dipole spectra for $^{130}$Sn
are\linebreak shown. The right part of the figure shows the dipole
strength function up to 26 MeV. The left panel shows the low-lying
parts of the corresponding spectrum below 12 MeV. Fi\-gu\-re
\ref{Sn130} (a) displays the experimental $B(E1)$ distribution.
The black circles with error bars are the experimental data. In
order to quantify this resonance-like structure the data are
fitted with a Lorentzian distribution \cite{akfb05,kpaf07}.
Calculations wit\-hin QRPA are shown in Fig. \ref{Sn130} (b), and
the QRPA plus phonon-phonon coupling (2PH) results are presented
in Fig. \ref{Sn130} (c). In the figure, the calculated $E1$
strength distributions are folded out with a Lo\-rent\-zian
distribution of 1~MeV width. The general shapes of the GDR
obtained in the 2PH are rather close to those observed in
experiments. This demonstrates the qua\-lity of a description
within 2PH in comparison with QRPA. We conclude that the main
mechanisms of the GDR formation in $^{130}$Sn are taken into
account correctly and consistently in the 2PH approach.

\begin{table}
\caption{Values of the centroid energy and width calculated within
the QRPA or QRPA plus phonon-phonon coupling (2PH) in comparison
with the experimental values (Expt.) taken from Ref.
\cite{akfb05}. The chosen energy interval is 11-20 MeV.}
\label{E1gdr}       
\resizebox{1.0\columnwidth}{!}{%
\begin{tabular}{llllllllll}
\hline\noalign{\smallskip}
 Nucleus   &\multicolumn{3}{l}{$\bar{E}$ (MeV)}&&\multicolumn{3}{l}{$\Gamma$ (MeV)}\\
\cline{2-4}\cline{6-8}
        & QRPA & 2PH & Expt.                  && QRPA & 2PH & Expt.   \\
\noalign{\smallskip}\hline\noalign{\smallskip}
 $^{124}$Sn& 16.4 & 16.3 & 15.3               && 4.4  & 4.7 & 4.8     \\
 $^{126}$Sn& 16.2 & 16.2 &                    && 4.4  & 4.7 &         \\
 $^{128}$Sn& 16.1 & 16.0 &                    && 4.7  & 4.7 &         \\
 $^{130}$Sn& 15.8 & 15.7 & 15.9(5)            && 4.8  & 4.8 & 4.8(1.7)\\
 $^{132}$Sn& 15.5 & 15.4 & 16.1(7)            && 4.9  & 5.0 & 4.7(2.1)\\
\noalign{\smallskip}\hline
\end{tabular}
}
\end{table}
\begin{table}
\caption{Mean energies $\bar E$ and summed $B(E1)$ values for the
low-energy dipole states in the excitation energy range below 11
MeV. The experimental values (Expt.) are taken from Refs.
\cite{akfb05,kpaf07,oe07}. To compare with experimental data in
$^{124}$Sn we choose the energy interval $E\leq 10$
MeV~\cite{oe07}.}
\label{E1pdr}       
\resizebox{1.0\columnwidth}{!}{%
\begin{tabular}{llllllllll}
\hline\noalign{\smallskip}
 Nucleus   &\multicolumn{3}{l}{$\bar E$ (MeV)}&&\multicolumn{3}{l}{$\sum B(E1)$ (e$^2$fm$^2$)}\\
\cline{2-4}\cline{6-8}
           &   QRPA   &   2PH   &   Expt.     &&  QRPA  &  2PH  &  Expt.  \\
\noalign{\smallskip}\hline\noalign{\smallskip}
 $^{124}$Sn&    9.7   &    9.1  &   6.97      &&  0.86  &  0.59 &  0.379(45)\\
 $^{126}$Sn&   10.1   &   10.0  &             &&  1.82  &  1.86 &         \\
 $^{128}$Sn&   10.0   &   10.0  &             &&  1.63  &  1.78 &         \\
 $^{130}$Sn&   10.0   &   10.0  &  10.1(7)    &&  1.40  &  1.80 &  2.4(7) \\
 $^{132}$Sn&    9.9   &    9.9  &   9.8(7)    &&  1.27  &  1.42 &  1.3(8) \\
\noalign{\smallskip}\hline
\end{tabular}
}
\end{table}

Let us discuss results for the low-energy $E1$ strength in more
detail (see in the left part of Fig. \ref{Sn130}). For the
low-lying part of the dipole spectrum, the QRPA calculation
predicts two rather pronounced states around 9.5 MeV. The neutron
transition densities of these levels are dominated outside the
nuclear surface. That corresponds to the vibrations of a neutron
skin against a proton-neutron core. One can see from Fig.
\ref{Sn130} (b) that the next fairly collective state is around
11.6 MeV. Using the analysis of the transition densities we found
that states above this energy have an intermediate behavior.
Increasing further the excitation energy we observe the low-energy
GDR tail. The corresponding 2PH strength distributions in Fig.
\ref{Sn130} (c) show many states with comparable strength in the
energy region below 11 MeV. The transition densities of these
states, originating the fragmentation of the QRPA pygmy mode, have
a behavior which is very similar to the behavior of the initial
QRPA state. However, we can state that the low-lying transition
densities are also dominated by the neutron contribution outside
the nuclear surface. Strictly speaking, a correct comparison of
the integrated pygmy dipole strength with experimental data in
this region is still an open problem, because of loss of
experimental sensitivity \cite{akfb05,kpaf07}.

Now we consider the integral characteristics of the\linebreak GDR.
The GDR calculated characteristics are displayed in Table
\ref{E1gdr}. They are compared with the experimental data
\cite{akfb05}. The centroid energy $\bar{E}$ and the spreading
width $\Gamma$ are given by the following expressions:
\begin{eqnarray}
\label{Ec} \bar{E}=\frac{m_1}{m_0}\,, \quad
\Gamma=2.35\sqrt{\frac{m_2}{m_0}-\left(\frac{m_1}{m_0}\right)^2}\,,
\end{eqnarray}
where $m_k=\sum_{}B(E1)\,{E}^k$ are the energy-weighted moments.
The integral characteristics of $E1$ strength function in these
nuclei have been calculated for the energy interval 11-20 MeV,
which exhausts between 100 and  110\% of the Tho\-mas-Reiche-Kuhn
energy-weighted sum rules ($14.8NZ/A$ MeV e$^2$fm$^2$).

First of all we checked our approach for $^{124}$Sn and obtained
(see Table \ref{E1gdr}) a reasonable agreement with available
experimental data \cite{akfb05}. In $^{124}$Sn we find that
$\bar{E}=16.4$ MeV for the QRPA and $\bar{E}=16.3$ MeV for the
2PH, even though the average energy of the giant dipole resonance
is by 1~MeV higher than its experimental value. This is in
agreement with the RRPA calculations \cite{pie06} where the
centroid energy for GDR is located at 15.86 MeV in the interval
10-25 MeV. For comparison, the QTBA \cite{agkk11} and RQTBA
\cite{lrtl09} calculations give 15.3 MeV and 15.05 MeV,
respectively. We note that in the QTBA the $0-30$ MeV summation
interval is used while in the RQTBA this interval is $10-22.5$
MeV. One can see from Table \ref{E1gdr} that the inclusion of the
two-phonon terms results in an increase of the resonance width
from 4.4 to 4.7 MeV. This value is in agreement with the QTBA
calculations (4.6 MeV) \cite{agkk11}.

The QRPA results for the neutron-rich $^{130}$Sn are presented in
Fig. \ref{Sn130} (b). The centroid value of the GDR is 15.8 MeV.
The results of our calculations for the $E1$ resonance taking into
account the two-phonon terms are shown in Fig. \ref{Sn130} (c).
Here, the GDR centroid energy is 15.7 MeV, that is in agreement
with another theoretical work using the RRPA \cite{pie06}
($\bar{E}=15.78$ MeV in the 10-25 MeV interval). The RQRPA and
RQTBA results are very similar ($\bar{E}=15.13$ MeV for RQRPA and
$\bar{E}=14.66$ MeV for RQTBA) in the interval 10-22.5 MeV
\cite{lrt08}. For the GDR width we obtain $\Gamma=4.8$ MeV that
can be compared with the other theoretical values 3.49 MeV in the
RQRPA and 4.74 MeV in the RQTBA. The experimental value of the
width is equal to 4.8(1.7) MeV \cite{agkk11}.

For $^{132}$Sn, values calculated by QRPA and 2PH for the GDR
energy centroid are $\bar{E}=15.5$ MeV and $\bar{E}=15.4$ MeV,
respectively. The experimental value of the GDR\linebreak width
for $^{132}$Sn is 4.7(2.1) MeV \cite{akfb05} and this is in
agreement with our results. The inclusion of the two-phonon terms
gives a small increase of the resonance width from 4.9 to 5.0~MeV.
Our results for $^{132}$Sn are qualitatively similar to
calculations in Ref.~\cite{sbc04} where the Skyrme interaction
SIII has been used. The peak energy and width with the phonon
coupling effects are 15.5 MeV and 5.8 MeV, respectively. It is
worth to note the GDR centroid energy in the interval 8-25 MeV is
14.7 MeV in the RRPA and it is 14.4 MeV in the RRPA with the
inclusion of particle-phonon coupling \cite{lrt07}. The inclusion
of particle-phonon coupling in the RRPA calculation results in an
increase of the resonance width from 3.3 to 4.0 MeV.

One can see that the agreement with the experimental data for
$^{124}$Sn and $^{130,132}$Sn is good. This gives us confidence to
use the model for a prediction of the integral characteristics of
the GDR in $^{126,128}$Sn.

In the case $^{126}$Sn, the centroid energy for the 2PH case has
the same value as for the QRPA calculation  16.2 MeV. One can see
that for $^{128}$Sn the centroid energy with 2PH is 16.0 MeV, to
be compared with the value 16.1 MeV of QRPA. The values of energy
centroids for $^{126}$Sn and $^{128}$Sn are rather close to the
empirical systematics $31.2A^{-1/3}+20.6A^{-1/6}$ \cite{bf75}. The
empirical predictions $120A^{-2/3}$ \cite{sw81} for the GDR widths
are reproduced well in both cases.

As it is seen from the Table \ref{E1gdr} our calculations show a
slight lowering of the GDR peak energy and an increase of the GDR
width when passing from $^{124}$Sn to $^{132}$Sn. These results
are in an agreement with the calculations of Litvinova
\cite{lrt08,lrtl09} and Piekarewicz \cite{pie06}. The role of
anharmonic effects, which mainly determine the value of the
GDR\linebreak width, is decreasing when passing from non-magic to
magic nuclei.

Finally, let us discuss the low-energy $E1$ strength. In contrast
to the GDR, the effects of phonon-phonon coupling is supposed to
be more serious. The integral characteristics of the strength
distribution of low-energy dipole states before 11 MeV are given
in Table \ref{E1pdr}. Note that to compare with experimental data
in $^{124}$Sn we choose the energy interval $E\leq 10$ MeV
\cite{oe07}.

In the case of $^{124}$Sn, the QRPA mean energy is equal to 9.7
MeV and taking into account phonon-phonon coupling gives rise to a
decrease of the PDR energy by 0.6 MeV, while experiment gives
$\bar E=6.97$ MeV \cite{oe07}. For comparison, the QTBA
\cite{agkk11} and RQTBA \cite{lrtl09} calculations give 8.7 MeV
and 8.15 MeV, respectively. The experimental data for the 4-10 MeV
interval give for the integrated PDR strength a value $\sum
B(E1)=0.379(45)$ e$^2$fm$^2$, while our calculations give values
0.86 and 0.59 e$^2$fm$^2$ within the QRPA and 2PH, respectively.
The calculated total QPM \cite{tl08} dipole strength in the PDR
energy range $E=5.7-7.2$ MeV is 0.324 e$^2$fm$^2$. This is rather
close to the experimentally deduced strength. On the other hand,
the QTBA \cite{agkk11} and RQTBA \cite{lrtl09} results are very
similar for the interval below 10 MeV ($\sum B(E1)=3.0$
e$^2$fm$^2$ within the QTBA and $\sum B(E1)=3.2$ e$^2$fm$^2$
within the RQTBA).

Values predicted by QRPA and 2PH for the PDR energy centroid in
$^{126}$Sn are  $\bar{E}=10.1$ MeV and $\bar{E}=10.0$ MeV,
respectively. The inclusion of the two-phonon terms results in a
slight increase of the summed $B(E1)$ from 1.82 to 1.86
e$^2$fm$^2$. In the case of $^{128}$Sn, the calculated summed
$B(E1)$ strength in the energy range below 11~MeV amounts to 1.63
e$^2$fm$^2$ (in the QRPA) and 1.78 e$^2$fm$^2$ (in the 2PH). The
centroid energy is 10.0 MeV in both cases.

Our calculations give for $^{130}$Sn a total dipole strength $\sum
B(E1)=1.40$ e$^2$fm$^2$ for the QRPA and $\sum B(E1)=1.80$
e$^2$fm$^2$ for the 2PH. The summation is performed for the dipole
states below 11 MeV. The experimental value is $\sum B(E1)=2.4(7)$
e$^2$fm$^2$ \cite{kpaf07}. On the other hand, the corresponding
mean energy of the PDR is 10 MeV (it is reproduced in both cases).
It may be compared with the experimental value 10.1(7) MeV. The
RQRPA and RQTBA results in the interval below 10 MeV \cite{lrt08}
are very similar $\sum B(E1)=4.04$ e$^2$fm$^2$ (RQRPA) and $\sum
B(E1)=3.44$ e$^2$fm$^2$ (RQTBA). The PDR centroid in the RQTBA is
equal to 7.5 MeV. For comparison, the RRPA calculations give
$\bar{E}=7.91$ MeV in the same interval \cite{pie06}. As one can
see from Figure \ref{Sn130} the inclusion of the two-phonon terms
results in an essential increase of the PDR width -
$\Gamma_{PDR}$. For the QRPA case we get $\Gamma_{PDR}=1.0$ MeV
that can be compared with the value $\Gamma_{PDR}=1.8$ MeV when
one takes into account the phonon-phonon coupling. Only an upper
experimental limit for the PDR width could be deduced because of
the finite energy resolution in experiment on Coulomb dissociation
\cite{akfb05}. This upper limit for the PDR widths is
$\Gamma_{PDR}<3.4$ MeV.

Now we discuss the low-lying dipole response in $^{132}$Sn for the
energy interval below 11~MeV. In RPA calculations the mean energy
is 9.9 MeV. Taking into account phonon-phonon coupling gives the
same value for the PDR energy, while experiment gives $\bar
E=9.8(7)$ MeV \cite{akfb05}. We notice that the inclusion of
phonon-phonon coupling in our calculations leads to an increase of
the $\sum B(E1)$ values from 1.27 to 1.42 e$^2$fm$^2$. In the QTBA
calculations \cite{agkk11} the centroid energy of PDR is located
at 8.9 MeV for the energy interval below 10 MeV. This value is by
0.9 MeV lower than its experimental one. For comparison, the RQTBA
\cite{lrtl09} calculations give 7.3 MeV for the same energy
interval. Our calculation shows that the inclusion of the
two-phonon terms results in an increase of the resonance width
from 1.2 to 2.0 MeV. For the widths of the PDR an upper limit is
$\Gamma_{PDR}<2.5$ MeV \cite{akfb05}.

Table \ref{E1pdr} demonstrates that there is a distinct difference
between characteristics of the low-lying strength integral in
$^{124-132}$Sn. Namely, the phonon-phonon coupling contribution to
$\sum B(E1)$ is small for nuclei in $^{126,128,130,132}$Sn, while
it is  rather important in the $^{124}$Sn. Moreover, for nucleus
such as $^{124}$Sn the PDR is almost completely defined by complex
configurations in the energy interval below 10 MeV.

The QRPA predicts a monotonic decrease of the PDR centroid with
mass number. The value of the total PDR strength decreases in
going to the heavier tin isotopes, up to $^{132}$Sn. Other QRPA
calculations with Skyrme forces \cite{tsnv07} give similar
tendencies for the summed $B(E1)$ values but our $\sum B(E1)$
values are somewhat smaller. The influence of coupling between
one- and two- phonon terms in the wave functions (\ref{wf2ph})
only leads to the fragmentation of $E1$ strength. On the other
hand, as one can see from Figure \ref{Sn130} the inclusion of the
two-phonon terms results in an essential increasing of the PDR
width. Also from the Table \ref{E1pdr} one can see that dipole
strength is transferred from the GDR region to the low-energy
region. In summary, our $\bar E$ and $\sum B(E1)$ values are in
good agreement with the experimental values, except for
$^{124}$Sn.

\section{Summary}
Starting from the Skyrme mean-field calculations, the properties
of the electric dipole strength in tin isotopes are studied by
taking into account the coupling between one- and two-phonons
terms in the wave functions of excited states. The finite rank
separable approach for the QRPA calculations enables one to reduce
considerably the dimensions of the matrices that must be inverted
to perform nuclear structure calculations in very large
configuration spaces.

Neutron excess  effects on the PDR excitation energy and
transition strength have been investigated for $^{124-132}$Sn. We
find that the centroid energy and summed strength decreases when
increasing the neutron number. The inclusion of the two-phonon
configurations results in an increase of the widths of the PDR and
GDR. This is in agreement with available experimental data. For
$^{126,128}$Sn the properties of the PDR are predicted.

\section*{Acknowledgments}
A.N.N., A.P.S. and V.V.V. thank the hospitality of IPN-Orsay where
a part of this work was done. The authors are thankful to
V.~Yu.~Pomonarev and E.~Khan for many fruitful and stimulating
discussions concerning various aspect of this work. This work was
partly supported by the IN2P3-JINR agreement No.~11-87 and the
RFBR No.~11-02-91054.

\end{document}